\documentstyle[prd,aps,floats]{revtex}
\begin{document}
\draft
\preprint{SUSSEX-AST 98/2-2, astro-ph/9802133}

%
%
\input epsf \renewcommand{\topfraction}{0.8} 
\twocolumn[\hsize\textwidth\columnwidth\hsize\csname 
@twocolumnfalse\endcsname

\title{The radiation--matter transition in Jordan--Brans--Dicke theory}
\author{Andrew R.~Liddle, Anupam Mazumdar and John D.~Barrow}
\address{Astronomy Centre, University of Sussex, Falmer, Brighton BN1
9QJ,~~~U.~K.}  
\date{\today} 
\maketitle
\begin{abstract}
We study the transition from radiation domination to matter domination in 
Jordan--Brans--Dicke theory, in particular examining how the Hubble length at 
equality depends on the coupling parameter $\omega$. We consider the 
prospects for using high-accuracy microwave anisotropy and large-scale 
structure data to constrain $\omega$ more strongly than by conventional solar 
system gravity experiments.
\end{abstract}

\pacs{PACS numbers: 98.80.Cq \hspace*{2.3cm} Sussex preprint SUSSEX-AST 
98/2-2, astro-ph/9802133}

\vskip2pc]


\section{Introduction}

One of the most important epochs in the history of the Universe is the
transition from radiation domination to matter domination. This transition 
alters the growth rate of density perturbations: during the
radiation era perturbations well inside the horizon are nearly frozen but
once matter domination commences, perturbations on all length scales are able
to grow by gravitational instability. Consequently, the horizon scale at the
time of matter--radiation equality is imprinted upon the spectrum of density
perturbations; indeed, in a flat cold dark matter cosmology it is the only
length scale to appear in the perturbation spectrum. The nature of this
transition has been well studied in the standard general relativistic
cosmology, and plays a crucial role in calculations of the perturbation
spectrum and associated microwave background anisotropies by codes such as 
{\sc cmbfast}~\cite{SZ}.

The success of general relativity as a description of our Universe allows us
to evaluate the performance of rival theories of gravitation. So far, no
weak-field or cosmological observations disagree with the predictions of
general relativity. The most important class of deviant theories are
scalar-tensor gravity theories, of which the Jordan--Brans--Dicke (JBD)
theory \cite{JBD,Will} is the simplest and best-studied generalization of 
general
relativity. This theory leads to variations in the Newtonian gravitation
``constant'' $G$, and introduces a new coupling constant $\omega$, with
general relativity recovered in the limit $1/\omega \rightarrow 0$. The most
robust constraint on $\omega$, that it must exceed 500, has been derived
from timing experiments using the Viking space probe \cite{Reas}, and has
stood for nearly 20 years now. Other constraints, such as those from
nucleosynthesis \cite{nucl}, are comparable but more model dependent; the 
most detailed analysis \cite{nucl2} gives only $\omega > 50$.

Within the next five to ten years, the advent of new microwave anisotropy
satellites, MAP and Planck, and large galaxy-redshift surveys, 2df and the
Sloan Digital Sky Survey (SDSS), promises to revolutionize our understanding
of cosmology by permitting the accurate determination of a large number of
cosmological parameters \cite{parest,Teg,parest2}. So far, estimates of the
accuracy of parameter estimation have only been made for cosmological
parameters, such as the Hubble constant $H_0$ and the matter density 
$\Omega_{{\rm m}}$, and for parameters describing the initial perturbations, 
such as the
spectral index $n$. However, such techniques can also in principle be
extended to constrain parameters defining the underlying gravity theory,
such as $\omega$, which also influence the gravitational instability
process. A full analysis of the viability of obtaining such constraints is
an imposing task; all the perturbation formalism employed to compute the
present-day matter and radiation power spectra must be generalized to the
theory of gravity under consideration, and the results then processed
through the Fisher information matrix technology of 
Refs.~\cite{parest,Teg,parest2}. In this paper, we assess whether such 
constraints might be competitive with existing bounds. We do this by studying 
the properties of the radiation--matter transition in JBD theory. We find 
that new cosmological data sets may well give limits competitive with those 
obtained from weak-field solar system tests of general relativity.

\section{The equations}

The equations for a zero-curvature Friedmann universe are 
\cite{JBD,Will,Wein}
\begin{eqnarray}
\left( \frac{\dot a}a\right) ^2+\frac{\dot a}a\,\frac{\dot \Phi }\Phi  
	&=& \frac{\omega}{6} \left( \frac{\dot \Phi }\Phi \right)^2
	+\frac{8\pi}{3\Phi} \rho \,;  \label{bdfield} \\
\ddot \Phi +3\frac{\dot a}a\,\dot \Phi  &=&\frac{8\pi }{2\omega +3}
	\left(\rho -3p\right) \,,  \label{bdfield2}
\end{eqnarray}
where the Brans--Dicke coupling, $\omega$, is a constant, $a(t)$ is the
cosmological scale factor, and $\Phi (t)$ is the Brans--Dicke field whose
present value gives the observed gravitational coupling. Here, $\rho $
and $p$ are the energy density and pressure of the cosmic fluid, which has
both matter ($p_{{\rm m}}=0$) and radiation ($p_{{\rm r}}=\rho _{{\rm r}}/3$) 
components, so $\rho =\rho _{{\rm m}}+\rho _{{\rm r}}$ and $p=p_{{\rm r}}$. 
Assuming negligible energy transfer between the fluids, they still obey
the general-relativistic fluid conservation equations with solutions 
\begin{equation}
\rho _{{\rm m}}=\rho _{{\rm m},0}\left( \frac{a_0}a\right) ^3\quad ;\quad
\rho _{{\rm r}}=\rho _{{\rm r},0}\left( \frac{a_0}a\right) ^4\,,
\end{equation}
where subscript `0' indicates the present value. We set $a_0 = 1$.

We shall assume a spatially-flat universe. It is important to realize that 
since the Brans--Dicke field appears in the Friedmann equation, the 
corresponding matter density will differ from the general relativity 
`critical' value, 
the correction being of order $1/\omega$, and this must be taken into account 
in our calculation. In fact, the present density of matter corresponding to a 
flat universe is higher in JBD theory, as the Brans--Dicke field contributes 
a negative effective energy density in the Friedmann equation.

The present radiation density has two components. The photon
energy density is dominated by the cosmic microwave background, whose
measured temperature $T_\gamma =2.728\pm 0.004$ K \cite{fixsen} gives 
\begin{equation}
\rho_{\gamma ,0} = 4.66 \times 10^{-34} \; {\rm g \, cm}^{-3}\,.
\end{equation}
However, the standard cosmology also contains relativistic
neutrinos. These cannot be directly detected, and so their contribution must
be fixed theoretically. Because neutrinos decouple before electron--positron
annihilation, they are at a lower temperature than the photons by a factor 
$\sqrt[3]{4/11}$ (see e.g.~Ref.~\cite{KT}), and, assuming three families of
massless neutrinos, their total contribution is $0.68$ times that of the
photons giving a present total density in relativistic particles of 
\begin{equation}
\rho_{{\rm r},0}=7.84 \times 10^{-34} \; {\rm g \, cm}^{-3} \,.
\end{equation}
Accurate modelling of neutrino decoupling may affect this at the percent
level, in a calculable way, as might contributions from unknown particles or
gravitons decoupling at much higher energies. The redshift of
matter--radiation equality is given by 
\begin{equation}
\label{equal}
1+z_{{\rm eq}}=\frac{\rho_{{\rm m},0}}{\rho_{{\rm r},0}}\\.   
\end{equation}
Our assumption of spatial flatness fixes the present matter density, where 
one must be careful to take the contribution of the Brans--Dicke field into 
account when defining the density parameter, $\Omega$.

\subsection{Analytic approximation}

When one of the fluid components dominates, the attractor solutions are well
known \cite{Nar}. For radiation domination, it is exactly the general
relativity solution 
\begin{equation}
\label{radsol}
a(t)\propto t^{1/2} \quad ; \quad \Phi ={\rm const} \,, 
\end{equation}
while for matter domination there is a slow variation of the gravitational
coupling described by the exact solution\footnote{In general, the JBD 
cosmologies have
exact solutions which show that they are dominated by the $\Phi$ field at
early times and by the perfect fluid matter fields at late times. The
general solutions approach the exact JBD vacuum solutions as $t\rightarrow 0$
and the particular fluid solutions (given by 
Eqs.~(\ref{matsol})--(\ref{matsol3}) in the dust
case) as $t\rightarrow \infty $ for a flat universe. These attractor
solutions also arise as simple exact solutions of Newtonian gravity with 
$G \propto t^{-n}$ and $a(t)\propto t^{(2-n)/3}$, see Ref.~\cite{attractor}.} 
\begin{eqnarray}
\label{matsol}
a(t) & = & \left( \frac{t}{t_0} \right)^{(2\omega +2)/(3\omega +4)} \,, \\
\label{matsolphi}
\Phi & = & \Phi_0\left( \frac{t}{t_0}\right)^{2/(3\omega +4)}\,, \\ 
\label{matsol1}
H_0 & = & \left( \frac{2+2\omega}{4+3\omega} \right) \, \frac{1}{t_0} \,, \\ 
\label{matsol2}
\rho = \frac{\rho_{{\rm m},0}}{a^3} & = & \left[ \frac{3H_0^2}{8\pi G} \, 
	\frac{(4+3\omega)(4+2\omega )}{6(1+\omega )^2}\right] \,
	\frac{1}{a^3} \,,\\  
\label{matsol3}
G & = & \left( \frac{2\omega +4}{2\omega +3}\right) \, \frac{1}{\Phi_0} \,.  
\end{eqnarray}
Here, Eq.~(\ref{matsol3}) relates the present measured value of the Newton
gravitation constant to the $\Phi$ field's value today, the relation being 
obtained by post-Newtonian expansion~\cite{Will}.
Eq.~(\ref{matsol2}) shows how the value of the matter density giving a 
spatially-flat universe is changed 
from the standard general relativistic expression by a finite $\omega$ value.

We are primarily interested in the Hubble radius at matter--radiation
equality, which sets the characteristic scale of the bend in the fluctuation
spectrum. It can be estimated analytically by assuming that the
matter-dominated solution holds all the way to equality; at equality, both
the radiation and matter contribute equally to $\rho$, of course. We solve
Eq.~(\ref{bdfield}) with the help of Eqs.~(\ref{matsol})--(\ref{matsol3}) to
estimate the Hubble radius. We take only the first-order dependence on 
$1/\omega $ in our calculation, which yields
\begin{equation}
\frac{a_{{\rm eq}}H_{{\rm eq}}}{a_0H_0}=\sqrt{2}\,\left( 
	\sqrt{1+z_{{\rm eq}}}\,\right)^{(2+\omega)/(1+\omega)}\times 
	\left[ 1+\frac{0.104}\omega \right] \,. 
\end{equation}

Using Eqs.~(\ref{equal}) and (\ref{matsol2}), we note that to $O(1/\omega)$
we have
\begin{equation}
1+z_{{\rm eq}}=24000h^2 \left(1+\frac 4{3\omega} \right) \,,
\end{equation}
where $h$ as usual is the present Hubble constant in units of $100\,{\rm 
km\,s}^{-1}\,{\rm Mpc}^{-1}$, leading to
\begin{equation}
\label{gert}
\frac{a_{{\rm eq}}H_{{\rm eq}}}{a_0H_0}=219h \times 
	\left[ 1+\frac{5.81}\omega+\frac{\ln h}\omega \right]    \,.
\end{equation}
The leading term in Eq.~(\ref{gert}) is the general-relativistic limit, and
is an exact result. Other terms in the equation are the corrections
accounting for the variation in $\Phi$ between matter--radiation equality and 
the present.

\subsection{Numerical approach}

The above result will be accurate at the large $\omega$ values that are of 
prime interest. However, to obtain the exact behaviour, we also tackle the
problem numerically.

In addition to the present values of $a$,
$\dot{a}$ and $\Phi$, we also require $\dot{\Phi}$ in order to specify a full 
solution of the JBD equations.
In practice, this additional freedom is eliminated by the
attractor behaviour during the long radiation-dominated phase. The attractor
solution can be picked out in the initial conditions by requiring that $\dot 
\Phi a^3$ approaches zero for $a$ tending to zero at the initial singularity 
\cite{JBD}. Integrating Eq.~(\ref{bdfield2}) once (taking advantage of the
radiation contribution cancelling out), we obtain 
\begin{equation}
\dot \Phi a^3=\frac{8\pi }{2\omega +3} \, \rho_{{\rm m},0} \, t + C\,,
\end{equation}
and choosing the growing mode amounts to setting the constant $C$ to be zero.

Our numerical code establishes the smooth transition between the radiation-
and matter-dominated solutions [Eqs.~(\ref{radsol}) and (\ref{matsol})]. We
study the Hubble length at the transition for all positive values of $\omega 
$ and plot our numerical output in Fig.~1, for two different values of $h$.
We compare our numerical result with the analytical estimate and 
confirm our analytical result at the asymptotic limit. At lower values of 
$\omega$, the analytic approximation becomes inaccurate (due to the neglect
of terms of order $\omega^{-2}$ and higher), and underestimates the change in 
$a_{{\rm eq}}H_{{\rm eq}}$.

\begin{figure}[t]
\centering
\leavevmode\epsfysize=6cm \epsfbox{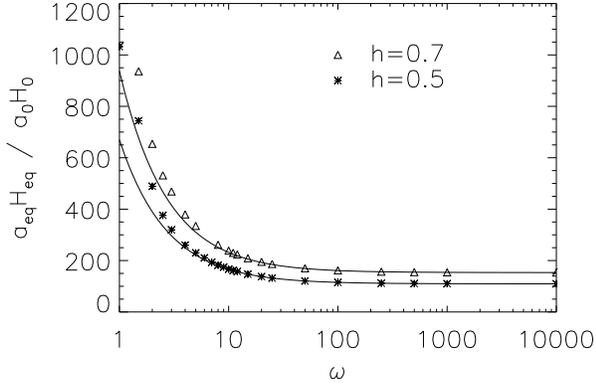}\\
\caption[fig1]{A comparison of numerical and analytical solutions of
the Hubble radius at equality for $h=0.5$ and $h=0.7$. The symbols correspond
to the numerical results, and the smooth curves to the analytical 
estimates of Eq.~(\ref{gert}) valid in the large $\omega$ limit.}
\end{figure}
	
\section{Parameter estimation}

The shift in matter--radiation equality will lead to a shift of the maximum
of the power spectrum. In the normal general relativity situation, and
assuming all dark matter is cold, this shift is governed by the combination
of $\Omega_{{\rm m}} h$, where $\Omega_{{\rm m}}$ is the matter density 
parameter, and this is often 
denoted $\Gamma$. Changing $\Gamma$
gives a horizontal shift to the power spectrum $P(k)$. Other parameters,
such as the baryon fraction of the total matter density, also influence the 
location
of the maximum, and sometimes $\Gamma $ is still used as an approximate
description of this \cite{Sug95}, although future observations are expected
to be so good that the baryons have to be included more accurately \cite{EH}. 
Other modifications to the cosmology, such as inclusion of a hot dark
matter component, cannot be modelled solely by a shift in the power spectrum.

At the current 95\% lower limit of $\omega = 500$, the shift in
matter--radiation equality will be at the one percent level. The parameter 
$\omega$ can also be probed via its affect on the growth rate of
perturbations although, because of biasing, galaxy surveys have problems in
constraining the overall normalization. Note that the lack of a
characteristic scale during matter domination means that the altered growth
rate will not change the shape of the spectrum; this result holds in JBD
theory as well as in general relativity.

Measurement of the shape of the power spectrum from galaxy surveys has been
considered in Refs.~\cite{Teg,parest2}. Tegmark \cite{Teg} introduces a
phenomenological parameter $\eta $ which shifts the power spectrum
horizontally, exactly the effect we are interested in, and considers how
accurately it can be measured by the SDSS. The accuracy depends on the
shortest scales considered; one cannot go too far without worrying about
non-linear clustering and biasing effects. If we take Tegmark's estimates
evaluated at the scale currently going non-linear ($k_{{\rm non-lin}}\simeq
0.1\,h \, {\rm Mpc}^{-1}$), we can read off the anticipated errors under two 
assumptions.
If all the model parameters are to be determined from the SDSS alone, 
$\Delta \eta /\eta \simeq 0.1$. If, on the other hand, it is assumed that all
parameters are already fixed except $\eta $, then $\Delta \eta /\eta \simeq
0.02$. This is entering at about the required level. At least in principle,
data on scales in the non-linear regime can improve this further if their
theoretical interpretation is deemed sufficiently unambiguous.

The case where other parameters can be considered as fixed might well apply
once one takes microwave background anisotropies into account, and uses them
to compute the other cosmological parameters. Indeed, if general relativity
is correct, then the Planck satellite alone will already be able to measure 
$\eta$ at the percent level \cite{Teg}, and then galaxy surveys can be used to
improve the estimate further. However, there are many potential
degeneracies: the horizon size at equality is changed by altering 
$\Omega_{{\rm m}}$, 
or $h$, or increasing the number of massless species (perhaps even by
including a possible thermal graviton background) as well as by introducing 
finite $\omega$. On the other hand, the degeneracy may be broken by the 
different 
growth rate of perturbations in JBD theory. The evolution of dust density
perturbations (where $\delta \equiv \delta \rho/\rho \ll 1)$ for all
wavelengths is determined by the solution of~\cite{naretal,Wein}
\begin{equation}
\label{pert}
\ddot \delta +2\frac{\dot a}{a} \, \dot{\delta} - \frac{8\pi \rho}{\Phi} 
	\, \left( \frac{2+\omega }{3+2\omega }\right) \delta = 0  \,.
\end{equation}
Hence, the growing mode solution of Eq.~(\ref{pert}) for the background 
Universe of Eqs.~(\ref{matsol}) and (\ref{matsolphi}) is given by 
\begin{equation}
\delta \propto t^{(4+2\omega )/(4+3\omega )}\,.
\end{equation}
This gives a normalization shift at the few percent level for $\omega \sim
500$, most of the effect being after the microwave anisotropies have been
generated, since the redshift of decoupling is close to that of equality.

It appears that there is a reasonable prospect that upcoming precision
observations can impose a limit on (or make a detection of) finite $\omega$
values at a level competitive with the post-Newtonian bounds, although the
use of galaxy surveys for this purpose is subject to a number of caveats
given in Ref.~\cite{Teg}. While we have presented the case for arriving at
new limits on the constant value of $\omega$ characterising JBD\ theories,
the basic technique can be extended to constrain the value of a non-constant 
$\omega(\Phi)$ defining cosmological solutions to a more general
scalar-tensor gravity theory, as discussed in Ref.~\cite{BPar}.

\section{Conclusions}

We have studied the matter--radiation transition in the JBD theory, both
numerically and analytically. The shift in the epoch of matter--radiation
equality will influence the shape of the density perturbation spectrum, and
it appears that precision microwave anisotropy measurements and
large galaxy-redshift surveys may in the future be able to impose limits on 
$\omega$ competitive with existing solar system bounds.

However, it seems unlikely that a very substantial improvement will be
possible. It may therefore be best to wait to see whether the high quality
of data promised is actually delivered before embarking on the substantial
undertaking of generalizing general-relativistic results to carry out a
proper estimate of the likely observational limits. If all goes well, use of
the data to constrain parameters of the gravitational theory will be a
worthwhile endeavour and an unexpected bonus from future high-precision
observational studies of galaxies and the microwave background.


\section*{Acknowledgments}

A.R.L. is supported by the Royal Society, A.M. by the Inlaks foundation and
an ORS award, and J.D.B. by PPARC. We thank Ian Grivell, Max Tegmark and
Diego Torres for useful discussions, and acknowledge use of the Starlink
computer system at the University of Sussex. 

\end{document}